# Results of experimental investigations of $^{60}$Co β-decay rate variation.

**Yu.A. Baurov[1], V.A. Nikitin[2], V.B. Dunin[2], N.A. Demchuk[3], A.Yu. Baurov[1], V.V. Tihomirov[2], S.V. Sergeev[2], A.Yu. Baurov (Jr)[1]**

[1]Closed Joint Stock Company Research Institute of Cosmic Physics, 141070, Moscow Region,
Korolyov, Pionerskaya, 4, Russia;
[2]Join Institute for Nuclear Reseach (JINR), 141980, Dubna, Russia
[3]Central Research Institute of Machine Building, 141070, Moscow Region,
Korolyov, Pionerskaya, 4, Russia

Results of long-term investigations of variation of $^{60}$Co β-decay rate from 28.12.2010 till 08.02.2012 are presented. The scintillation spectrometer with two LaBr$_3$ detectors is used to register of γ-quanta with energy 1.173 and 1.332 MeV accompanying $^{60}$Co β-decay. Counting rate of each detector and their γ-quanta coincidence are collected in successive time intervals 10 s. The statistical Kolmogorov-Smirnov method for data analysis is used. Temperature influence on experimental results is also analyzed. Deviations of β-decay counting rate from constant distribution during the days were detected in those decades: from 11.03 to 21.03 with significance level a = 0.1; from 22.04 to 02.05 with a=0.0125; from 24.06 to 04.07 with a=0.05; from 04.08 to 14.08 with a=0.05.

## 1. Introduction

In classic nuclear physics the β – and α – decays of radioactive elements are usually thought to be purely random processes practically not subjected to exterior influences [1, 2]. In recent years the papers [3 - 14] appeared in which the changes in the decay rates of various radioactive elements above the measuring errors are recorded. For example, a non-exponential radioactive decay of nuclei-isomers of $^{125m}$Te during uninterrupted measurements of γ-radiation in the course of a year was observed [4]. These changes are shown to have a periodic character [5 – 14]. In the paper [5] annual oscillations of the half-period $T_{1/2}$ of $^{32}$Si (β-decay) are found out. This study was performed in the Brookhaven National Laboratory (BNL, USA) from 1982 till 1986. The results of $T_{1/2}$ measurements carried out in Germany at the "Physicalisch-Technische Bundesanstalt (PTB)" are presented in [6]. The ionization chamber for $^{226}$Ra α – decay detection is used. Note that the paper [15] criticizes studies [5, 6] in the context of possible influence of the thermal factor on equipment performance.

In Russia yearly [7], near-daily and 27-day oscillation [7 – 14] were detected by various teams of researchers. It is worth to clarify that [7] does not consider variations of the decay products flow, but only repetitive duplication of the spectrum shapes and their deviation from the Poisson distribution. Ref. [16] contains critical remarks on the paper [7]. It was also shown that the variation of the β-decay rate of radioactive elements when laboratory moves together with the Earth rotation, reveals three distinct directions in the physical space relative to stationary stars [8 - 14].

Ref. [17] presents preliminary results of the experimental study of the $^{60}$Co β-decay rate in period February-August 2010 that confirms the presence of the two of the above mentioned three spatial directions [10–14].

In 2010 the experimental installation was being debugged. In particular the influence of the temperature on the results was investigated. The analysis of the [17] paper results has shown the necessity of placement of the experimental installation into the thermostat with high level of the temperature stability, while using scintillation detectors.

The present paper is a continuation of the investigations initiated in Ref. [8 – 14] and [17]. The purpose is the experimental method improvement without considerable expenses and to minimize systematic errors at registration of $^{60}$Co β-decay rate variation within a day or longer intervals.

## 2. The experimental installation and choice of β source

The choice of radioactive source is determined by the results of previous experiments and the intention to maximize the effect of action of the hypothetical new force of the nature responsible for the variation in the β-decay rate of radioactive elements and to minimize systematic errors. The analysis of experimental results [8 – 14] has shown that the value of the β-decay rate variation is larger for the radioactive elements having the nucleus with the large magnetic moment. This condition corresponds with the nature and feature of the new force. Thereupon the radioactive β source $^{60}$Co was chosen for the investigation. The presence of two lines 1.117 MeV and 1.332 MeV in its spectrum allows the two γ-quanta detection in coincidence and minimize electronics-dependent systematic errors.

Let's shortly describe experimental installation. It is placed in the thermostat that keeps temperature within the limits $30.6^{\circ}C \pm 0.1^{\circ}C$. The detecting module contains a radioactive β source $^{60}$Co and two-arm scintillation γ-spectrometer in hermetic case. Either of the two detectors records about 32 thousand of γ-quanta per second. LaBr$_3$(Ce) crystal is used as scintillation counter, relative energy resolution of which equals 3%. The crystal has diameter 25 mm and height 12 mm. The flowchart of the γ-spectrometer electronics is shown in Fig.1. It consists of two identical microprocessor-based assemblages and common control and power supply blocks. To receive the synchronization signal, the apparatus has an interface CAN on the motherboard.

The light pulses generated in the active zone of scintillation detectors are transformed into electron pulses by photomultipliers PM-1 and PM-2. The signals from PMs are additionally amplified and shaped by amplifies-shapers AF-1 and AF-2. The formed pulses are further transmitted to the discriminator units DU2-1, DU2-2, and two microprocessor assemblies containing a microcontroller MC, discriminator units DU5, and programmable logic integrated circuit (PLIC). Each discriminator unit DU2-1, DU2-2 contains two discriminators which thresholds are adjusted for detection of the pulse amplitudes in the range from 0.3 MeV to 3 MeV.

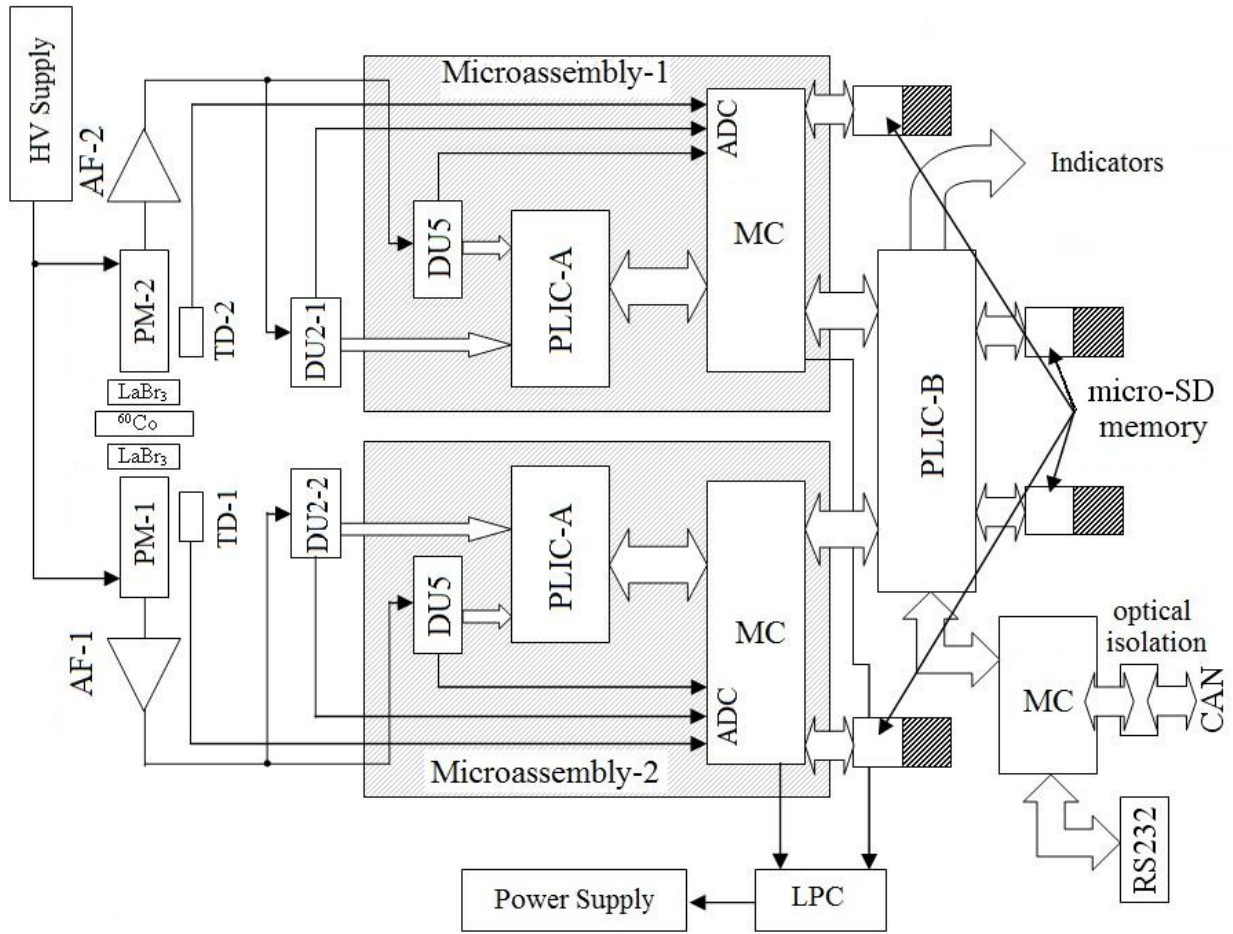

**Figure 1.** Gamma-spectrometer flowchart.

HV – High Voltage, PM – photomultiplier, AF - amplifier-shaper, TD – thermal detector, DU - discriminator unit, PLIC - programmable logic integrated circuit, ADC – analog-digital convertor, MC – microcontroller, LPC - power supply latching circuit.

The coincident pulses in both channels are recorded by counters. The reference voltages of the discriminators come to the microcontroller ADC and get stored there. The discriminator unit DU5 contains five discriminators adjusted for pulses registration in the energy ranges >0.3 MeV, > 1.1 MeV, > 1.3 MeV, > 1.6 MeV, and 2.8 MeV for further spectral analysis. The reference voltages of this discriminator are transmitted to the microcontroller ADC.

The pulses count is summarized in each 10 seconds interval, then read out by microcontroller and recorded into the external memory realized on microdrives that allow data recording onto replaceable 2 GB memory cards. The temperature is also registered at several specially separated points of the apparatus and recorded. Information is read out from the memory cards within the time interval from 10 days to 3 months. Characteristic γ-ray radiation spectrum is shown in Figure 2. It should be noted that the spectrometer energy calibration can be easily done. Three lines 1.173 MeV, 1.332 MeV and 1.41 MeV are clearly separated in the spectrum. The latter belongs to $^{40}$K natural isotope impurity. Around the energy range ≈ 2.5 MeV there is an enhancement caused by overlapping of the signals from natural impurity of heavy radioactive elements.

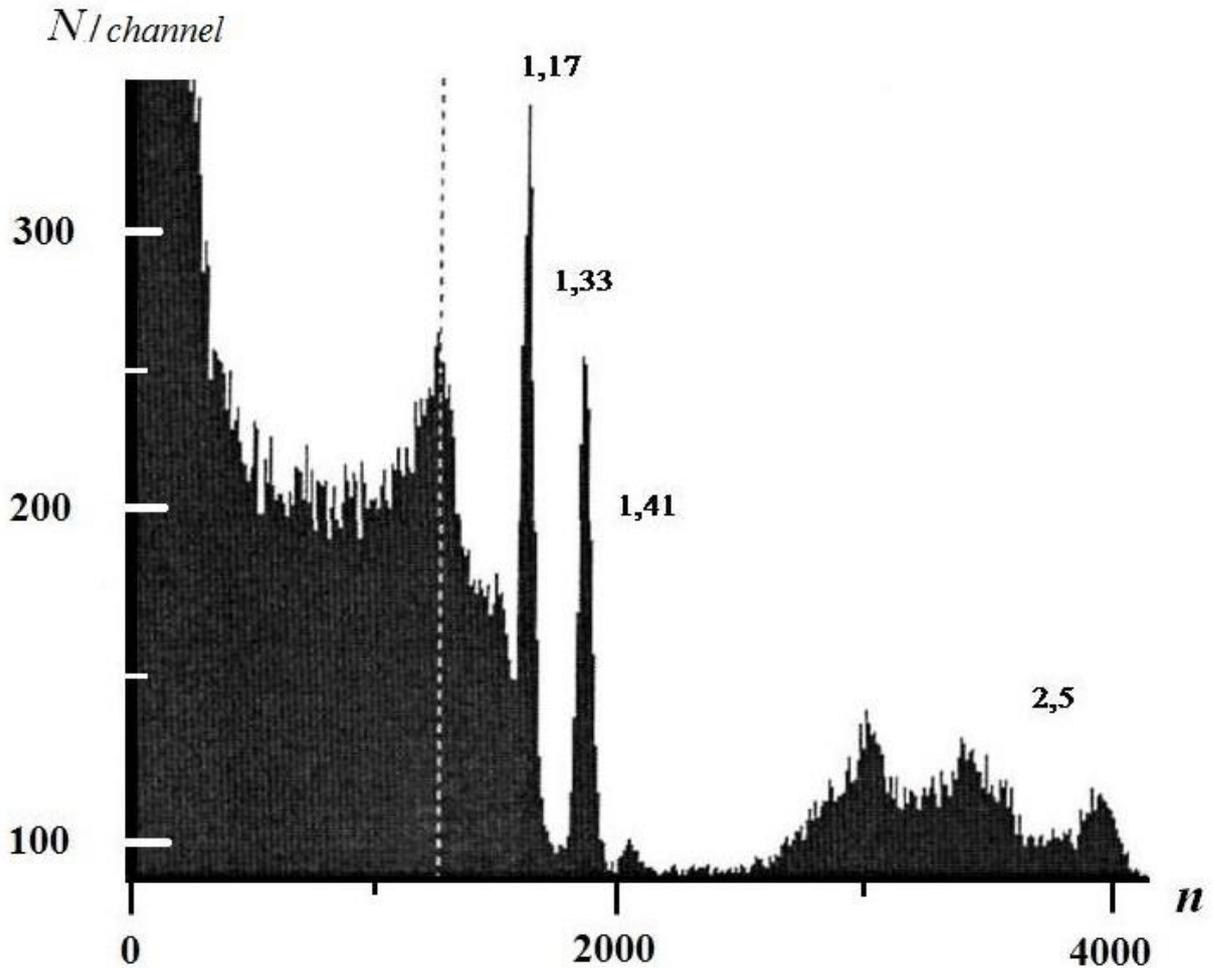

**Figure 2.** $^{60}$Co γ-ray radiation spectrum, recorded by use of LaBr$_3$(Ce) scintillation counter.

The number of γ-quanta per channel (*N*/channel) is plotted against of the ordinates axis and the number of *n-th* channel - against of the abscissas axis. Characteristic lines of the $^{60}$Co (1.17MeV and 1.33MeV), natural isotope impurity $^{40}$K (1.41MeV) and impurity of heavy radioactive elements (2.5MeV) are indicated with numbers.

These signals does not interfere normal running of the experiment, moreover it can be used for spectrometer calibration.

## 2. The results of the experiment

The present paper contains results of the experiment on the search of deviations in $^{60}$Co β-decay from the standard exponent. The exposition covers period from December 28, 2010 to February 08, 2012. Figure 3 shows daily average results for γ-quanta flows, registered by the first channel with the threshold 0.3 MeV, as well as values of temperature in relative units. The second channel gives the similar results. Discrete steps in diagram (1) and temperature (2) variations in the first half of the picture correspond to replacement of the installation memory cards.

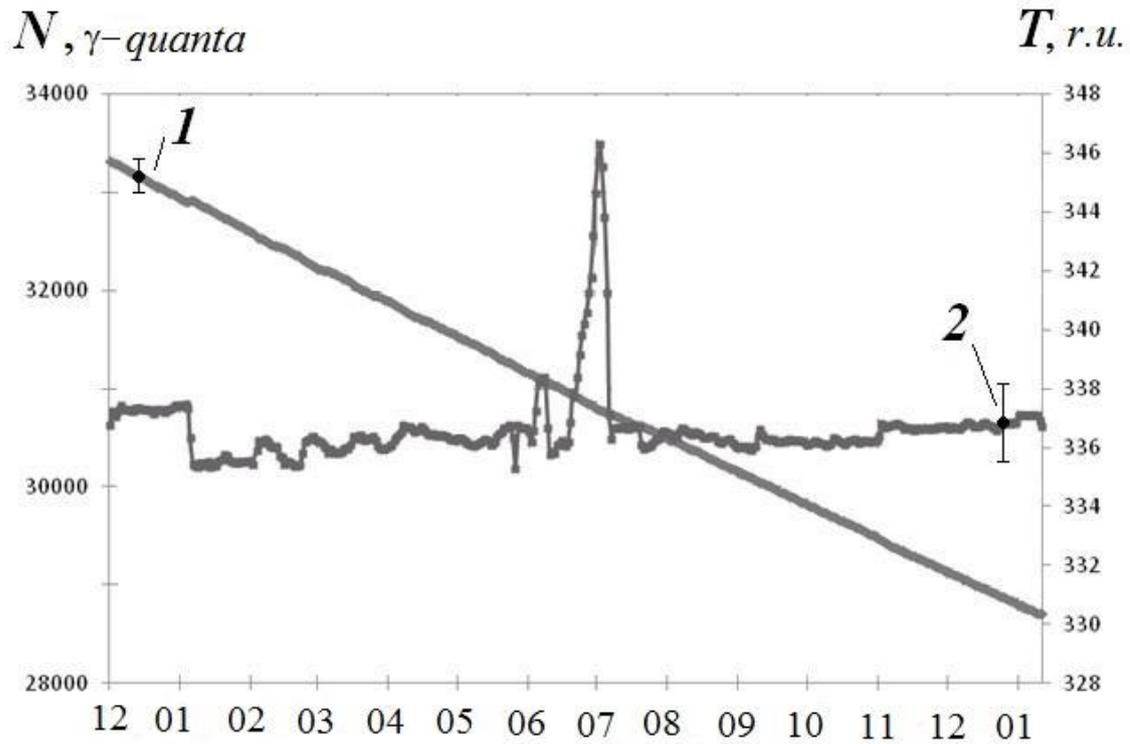

**Figure 3**. Daily average *N* - γ-quanta count (*1*) with the threshold > 0.3 Mev for the first channel and *T* - daily mean temperature (*2*) in relative units (r.u.) for the exposition period December 28, 2010 - February 08, 2012 (point on abscises axis – the 28th of every month).

Abrupt change of the temperature (at a level of 2.7%) at the end of July was caused by the thermostat failure to maintain temperature due to abnormal hot weather in Moscow that time. As consequence the temperature in thermostat exceeded of the premise level (30.6°C ± 0.1°C). Temperature variations level (except for the above abnormality and short periods of memory cards removal) does not exceed 0.2°. The data analysis reveals correlation between the intensity of γ-quanta and the temperature value. The corresponding small corrections are applied to the counting value and have not influence on the results.

Two methods are used to analyze the data: Fourier data sequence analysis method and Kolmogorov-Smirnov (K-S) statistical method [9, 11–14]. K–S criterion is based on computation of maximum difference between the theoretical hypothesis and the experimental distribution function. This difference is compared with tabulated Kolmogorov function. In our case the theoretical function is the uniform distribution of γ-quanta intensity in the course of a day. The criterion value for two decades of continuous data collection is shown in Figures 4 and 5 on the vertical (K-S) axis. The horizontal line with the significance level *a*, $a = 0.0125$ (Fig. 4) and $a = 0.3$ (Fig.5) corresponds to the criterion compliance probability. If K–S goes above the horizontal line, the hypothesis is rejected with probability *1-a*, for example, 0.9875 for Fig. 4, therefore there is statistically significant difference between the experimental and theoretical distributions. In the case under consideration, where theoretical distribution is selected to be uniform, this means nonuniformity of the measured value within the course of a day.

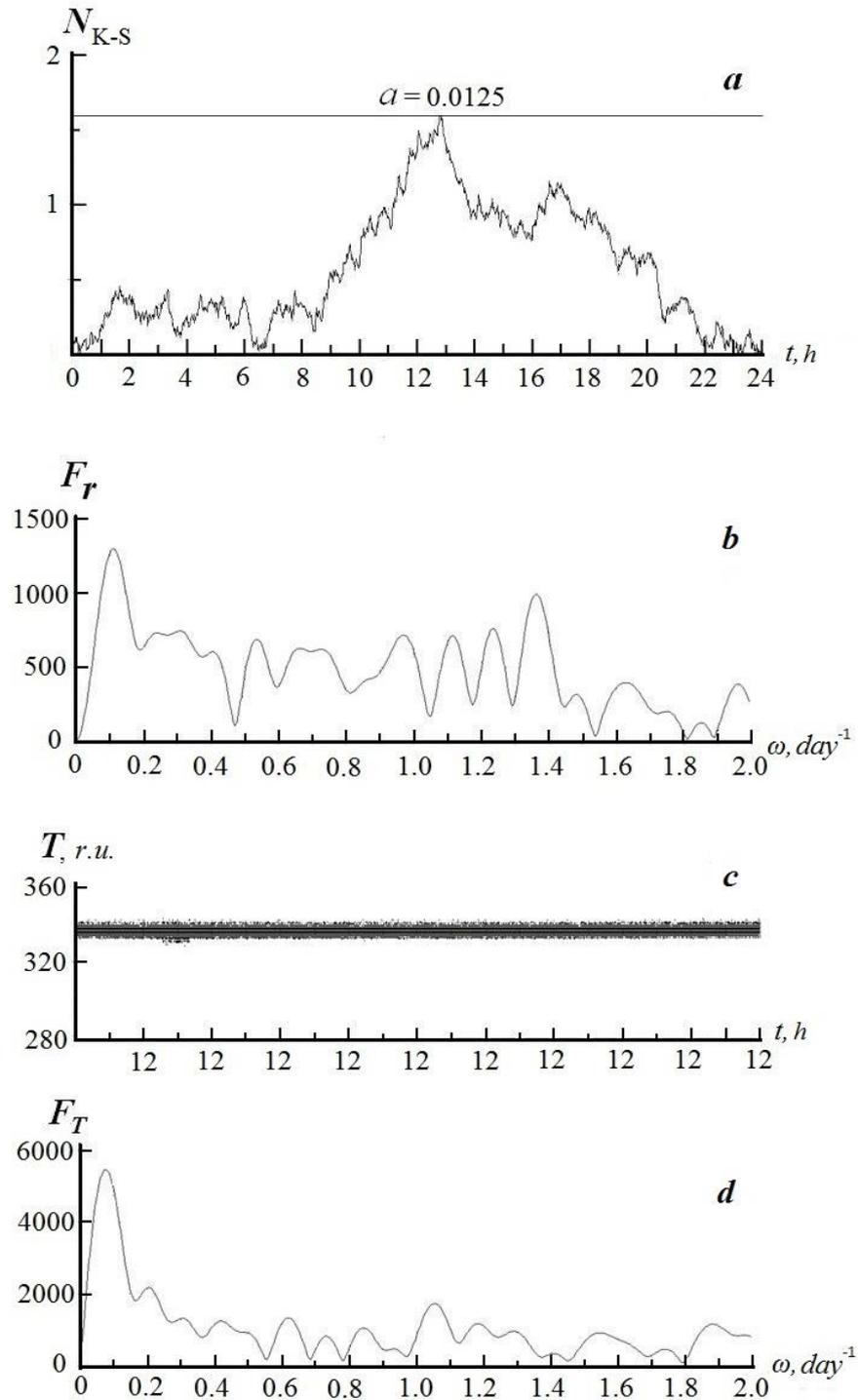

**Figure 4.** Data analysis for the decade April, 22 - May, 02, 2011.

Gamma counting rate registered in coincidence: *a* – nonuniformity assessment of γ-quanta counting rate distribution by K-S criterion. For analysis selected time intervals where counting rate exceeds mean value on 2 standard deviations (2σ). $N_{K-S}$ criterion values are plotted against the ordinate axis. The horizontal line corresponds to $N_{K-S}$ criterion value with the significance level $a = 0.0125$. The data cover the 24-hour interval (the Moscow time *t*, hours, is plotted against the abscissa axis. *b* - Fourier spectrum $F_R$ for γ-quanta counting rate, exceeding 2σ; *c* - the temperature series *T* values in relative units; *d* - Fourier spectrum $F_T$ for the temperature series.

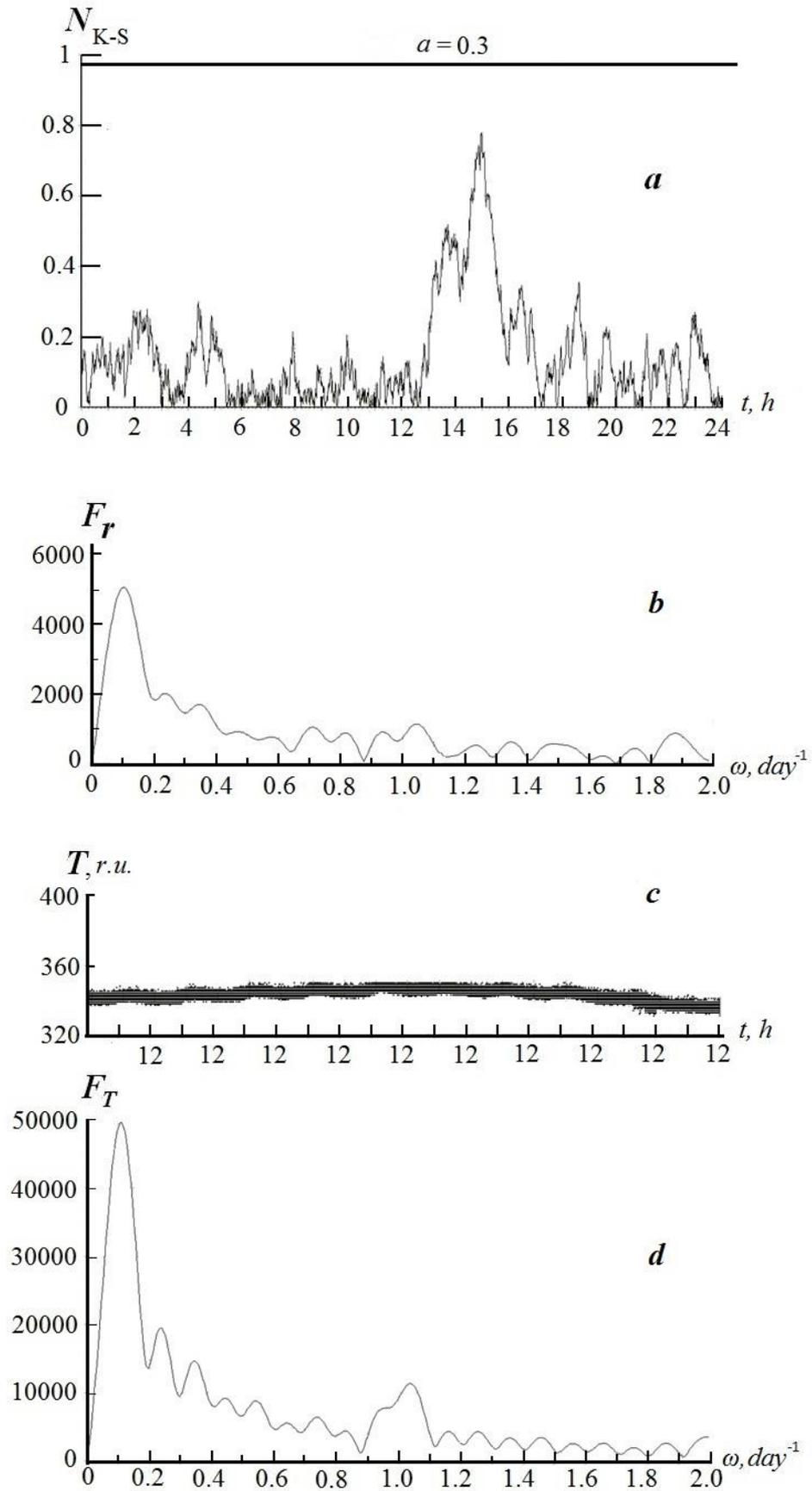

**Figure 5**. Data analysis for the decade July 23 - August 02, 2011.
All designations are the same as in the Figure 4.

For the present study we selected the periods of 10 days. Intensity in the γ-γ coincidence channel is under investigation. We are interested in a daily variation of γ-γ counting rate. To increase statistical accuracy within 24-hour period we add data of 10 day intervals in succession, i.e. the first bin (count in 10 seconds interval) of the first day of the first selected interval is matched with the first bin of the second interval and corresponding data are summed. The same summation procedure is made for the rest of the days of selected intervals. There is another-peculiarity of the present analysis. From the reference series only those bins are selected in which the counting rate exceeds the local average value for 2σ, where σ - is a statistical error in of the given bin. The series obtained in this way are listed below.

Assessment of the compatibility of observed series of count with uniform distribution by K–S test is shown only four decades with the significance level *a* less than 0.3:

March 11-21, with the significance level *a* = 0.1; temperature correlation ~ 0.05%;

April 22 - May 02, with the significance level *a*=0.0125; temperature correlation ~ - 0.04%;

June 24 - July 04, with the significance level *a*=0.05; temperature correlation ~ 2.1%;

August 04 - 14, with the significance level *a*=0.05; temperature correlation ~ 0.2%.

As an example, figure 4 shows the data analysis for the decade April 22 - May 02, 2011. As it is seen in the figure, the significance level of the result (non-uniformity of the decay rate over the 24-hour period) in its maximum reaches 0.0125. Fourier analysis of the 24 hours harmonics also indicates the process non-uniformity. Thus harmonics with the highest power for the main and temperature series are not coincide. The figure shows temperature stability in relative units (r.u.) without hourly harmonics (deviation ± 1.5%). Figure 5 shows the data analysis for the decade July 23 - August 02, 2011, where the temperature perturbation was observed (Fig. 3). It caused by abnormal hot weather in Moscow that time. The thermostat fails to maintain preset temperature. As figure shows, the daily temperature harmonic practically does not influence the rise of the power of cophasal harmonic in the γ-quanta spectrum. Thus K–S criterion value does not reach the limiting significance level of 0.3.

Now consider possible long-periodic harmonics in $^{60}$Co decay rate. A classical exponential decay with half-life $T_{1/2}$ = 5.2±0.3 year is observed, that corresponds to the tabular value 5.2713(8) year [18]. Measured values $T_{1/2}$, obtained by decades (selected decades with minimal temperature correla-tion) for the period December 28, 2010 - February 08, 2012, are shown in Figure 6. Some decades show deviations from the tabular value $T_{1/2}$, but they are not beyond the measurement accuracy range. 10-days and 20-days periods in decay rate change were detected by applying Fourier analysis, which seems to be an artefact caused by memory cards replacements. Additionally the data indicate the existence of 27-day period.

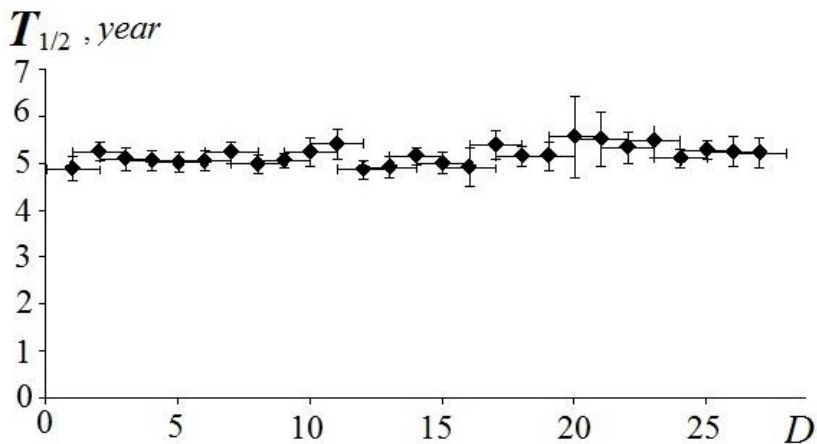

**Figure 6**. The half-decay time $T_{1/2}$ of $^{60}$Co measured in 10 day intervals. The periods of time are selected when temperature variation is small. $T_{1/2}$ values are plotted against of the ordinates axis. Decades (*D*) are plotted against of the abscissa axis. The standard deviations are shown.

### 3. Discussion

Standard representations can't help in explaining the obtained results - $^{60}$Co decay rate changes. Therefore it indicates existence of a new force in nature that influences neutron decay in $^{60}$Co nucleus.

The physics of hypothetical new force is described in [12 - 14]. Its influence on the β-decay is discussed also in [10, 17]. Let us briefly elucidate its nature. We work in the framework of a non-calibration theory. We elucidate formation of physical space and world of elementary particles. A part of mass of elementary particle associated with the process of formation of its internal physical space is proportional to the modulus of some summary potential $A_\Sigma$ [12-14]. The magnitude of $A_\Sigma$ modulus cannot exceed the modulus of the cosmological vector potential $|\mathbf{A}_g|$. A new fundamental constant $|\mathbf{A_g}| = 1.9 \cdot 10^5 \text{ T·m}$ is introduced for the first time in [19]. If the magnitude of $A_\Sigma$ is diminished by any way, the particle will undergo the action of new force throwing out substance from the region with diminished modulus of $A_\Sigma$. In our experiment the magnitude of $A_\Sigma$ decreases due to the action of the Earth's magnetic field vector potential $A_E$ which is practically tangent to the Earth's parallel. If vector $A_E$ is antiparallel to the vector $A_g$ then as a result there is decrease of the vector potential of the Sun's magnetic field dipole component $A_s$ entering into $A_\Sigma$. In April 2011 $A_s$ is directed like $A_g$, and in October it is in antiparallel to $A_g$. Therefore in April the magnitude of $A_s$ is diminished and hence $A_\Sigma$ is diminished also due to $A_E$. But this is not a case in the second half of year (from middle of July till middle of January) since in that time the vector $A_s$ has no components directed parallel to $A_g$ but only opposite to $A_g$. It should be accentuated once more that in this case particle masses cannot be decreased by the action of vector $A_E$. In this connection we see the maximum effect in April (for given solar cycle).

Obtaining of a result in one decade of August, 2011 (04.08–14.08) and its absence in the same decade in 2010 is indicative of the approaching change in dipole component polarity of the Sun's magnetic field, which occurs every 11 years. For the last time the similar process took place in 1999-2000 and lasted about 280 days [20]. After the Sun's magnetic dipole polarity is changed the maximum effect is to be observed in October.

So this experiment confirmed some previously obtained results. It also conform to anisotropic properties of the physical space described in [14, 21] and based on the observation of annual oscillations of $^{32}$Si and $^{226}$Ra half-decay periods [5, 6], being actively discussed in the references [15, 22, 23].

The number of papers devoted to the investigation of decay rate variation of radioactive elements increase [24, 25].

The present analysis uses only part of obtained information, therefore the quoted results to be treated as preliminary ones. An indication of the variation of $^{60}$Co isotope decay time was obtained. It is important to confirm this effect in course of future more comprehensive analysis of the data available and continuing measurements. It is important to determine such variation characteristics in detail: its amplitude and specific time dependence obtain more precise astronomic time reference, study out solar activity dependence. The equipment created is also a subject to some modifications. In particular, the counting rate can be increased by at least one order of magnitude.

REFERENCES

1. Y. M. Shirokov, N. P. Yudin, Nuclear Physics, Second edition, Moscow, "Nauka", 1980 ,(in Russian).

2. D. N. Trifonov, Radioactivity Yesterday, Today, Tomorrow, Atomizdat, Moscow, 1966, p. 91, (in Russian).

3. T. Ohtsuki et al., Phys. Rev. Lett. **93,** 112501 (2004).

4. S. K. Godovikov, Pis'ma Zh. Erksp. Teor. Fiz. **79**, 249; (2004) [JETP Lett. **79**, 196 (2004)].

5. D. E. Alburder, G. Harbottle, and E. F. Norton, Earth and Planet. Sci. Lett. **78,** 169 (1986).

6. H. Siegert, H. Schrader, and U. Schotzig, Appl. Radiat. Isot. **49,** 1397 (1998).

7. S. E. Shnol et al. Uspehi phys. Nauk, **168**, 1129 (1998), (in Russian).

8. Yu. A. Baurov and V. L. Shutov, Prikladnaya Fizika, No. 1, 40 (1995)., (in Russian)

9. Yu. A. Baurov, A. A. Konradov, V. F. Kushniruk and Yu. G. Sobolev, Scientific Report № E7-97-206, (Dubna, 1997), p.354.

10. Yu. A. Baurov, A. A. Konradov, E. A. Kuznetsov et al., Mod. Phys. Lett. A **16**, 2089 (2001).

11. Yu. A. Baurov, Yu. G. Sobolev, Yu. V. Ryabov, V. F. Kushniruk., Phys. Atom. Nucl. **70,** 1825 (2007).

12. Yu. A. Baurov, On the structure of physical vacuum and a new interaction in Nature (Theory, Experiment and Applications) (Nova Science, NY, 2000).

13. Yu. A. Baurov, Global Anisotropy of Physical Space. Experimental and Theoretical Basis (Nova Science, NY, 2004).

14. Yu. A. Baurov, Bull. Russ. Acad. Sci. Phys. **76,** 1076 (2012).

15. J. C. Hardy, J. R. Goodwin and V. E. Iacob, Arxiv: 1108.5326.

16. A. V. Derbin et al., Uspehi phys. Nauk, **170,** 209 (2000) (in Russian).

17. Yu. A.Baurov, N. A.Demchuk, A. Yu.Baurov, A. Yu.Baurov (Jr.), V. B. Dunin, V. V. Tihomirov, S. V. Sergeev, Prikladnaya fizika **5**, 12 (2011).

18. G. Audi, O. Bersillon, J. Blachot and A. H. Wapstra, Nuclear Physics A **729,** 3 (2003)

19. Yu. A. Baurov, Yu. N. Babaev, V. K. Ablekov, Sov. Phys. Dokl. **26**, 733 (1981).

20. A. N. Zhukov, I. S. Veselovsky, Proc. I[st] Solar Space Weather Euroconference. " The Solar Cucle and Terrestrial Climate", Santa Cruz de Tenerife. Tenerife. Spain. 25-29 September2000 (ESA SP-463, December 2000), p. 467.

21. Yu. A. Baurov, I. F. Malov, Int. J. Pure & Appl. Phys. **6**, 469 (2010).

22. J. H. Jenkins et. al. Astro-ph/0808.3283, 25 Aug. (2008).

23. E. Fischbach et. al. Space Sci. Rev. **145,** 285 (2009).

24. Bellotti et.al. Phys. Lett. B,**710,** 114 (2012).

25. J. H. Jenkins, E. Fischbach, D. JavorsekII, Robert H. Lee, P. A. Sturrock, Appl. Radiat. Isot. **74,** 50 (2013).